# Agent-based Visualization of Streaming Text

Jordan Riley Benson, Phillip Lafleur, David Crist and Benjamin Watson
North Carolina State University

**ABSTRACT**

We present a visualization infrastructure that maps data elements to agents, which have behaviors parameterized by those elements. Dynamic visualizations emerge as the agents change position, alter appearance and respond to one other. Agents move to minimize the difference between displayed agent-to-agent distances, and an input matrix of ideal distances. Our current application is visualization of streaming text. Each agent represents a significant word, visualizing it by displaying the word itself, centered in a circle sized by the frequency of word occurrence. We derive the ideal distance matrix from word co-occurrence, mapping higher co-occurrence to lower distance. To depict co-occurrence in its textual context, the ratio of intersection to circle area approximates the ratio of word co-occurrence to frequency. A networked backend process gathers articles from news feeds, blogs, Digg or Twitter, exploiting online search APIs to focus on user-chosen topics. Resulting visuals reveal the primary topics in text streams as clusters, with agent-based layout moving without instability as data streams change dynamically.

## 1 INTRODUCTION

Human exposure to information far exceeds our ability to manage it [3]. Constant streams of data from 24-hour news sources, blogs, and social networking sites bombard us and have created a market for information management tools, including the monitors described in [6]. Yet most of these tools fail in the face of today's rapidly increasing information throughputs, falling hopelessly behind current trends, or missing important events altogether. Still others do not support effective customization and tuning to applied needs or the local work environment.

Ishizaki [5] was one of the first to recognize the need for dynamic visualization of data streams, and designed an agent-based system to visualize news feeds. However, it was ahead of its time, dealing with low throughputs and just a few information sources. TextPool [1] was a more recent attempt, using many of the same information retrieval techniques we describe here, but required expert tuning to avoid instability in spring-mass layout, could not make use of online filtering APIs, and restricted itself to displaying absolute co-occurrence. The Buzz [4] permits users to choose among multiple information types and visualizes them with attractively, but performs little analysis of the data stream.

Our infrastructure draws on the strengths of these systems and extends them. Its use of agents enables dynamic and robust response to today's very high and unpredictable information bandwidths. Users can tune agent behaviors to support continuous passive viewing on the desktop, and intermittent active or passive use in a public space. The streaming text visualization application built with this infrastructure displays an unusually deep analysis of the data stream, showing word co-occurrence in the context of word frequency. Finally, the infrastructure makes effective use of today's Web 2.0 APIs, enabling the use of a wide range of data sources, and permitting users to filter those sources to meet their interests and applied needs.

Dept. Computer Science, 890 Oval Dr., Box 8206, EBII, Raleigh NC 27695-8206. bwatson@ncsu.edu

## 2 AGENT-BASED VISUALIZATION INFRASTRUCTURE

In our infrastructure, agents connect individual data elements (e.g. points, records, words) to visual appearance and position. Agents update themselves dynamically in response to changes in streaming data, using behaviors passed from the application.

Agents control their position by referencing two $n \times n$ distance matrices, where $n$ is the number of data elements and agents. The application regularly updates the ideal distance matrix $D_i$, which describes the ideal distances between all pairs of data elements. The second displayed distance matrix $D_d$ describes the distances between all pairs of agents in the current display. During each iteration, we derive the matrix $\Delta D = D_i - D_d$. Each entry $\Delta D[i,j]$ describes the attraction of agent $i$ to agent $j$. Each agent $i$ then moves with velocity $v_i$ where

$$v_i = \frac{\sum_{j=1}^{n} \Delta D[i,j] \overline{(p_i - p_j)}}{n}$$

and $p_i$ is the 2D position vector of agent $i$. An arbitrary coefficient can be applied to increase or decrease movement speed as desired.

This distance-based method of position control provides a useful interface for depicting high-dimensional data of all sorts. Applications may also pass agents alternative or additional behaviors such as collision avoidance and preventing motion outside of display space.

## 3 STREAMING TEXT VISUALIZATION APPLICATION

Our text stream monitoring application is fed by news aggregators, blogs, Twitter, Digg and RSS feeds. It has three components: the visualizer, the scraper, and the analyzer.

The *visualizer* maps agents to individual words, displaying them with the word itself, embedded in a circle sized to depict word frequency. Circles redden when word position is less ideal, meaning $\sum \Delta D[ij]$ is non-zero.

To cluster related words, the visualizer fills the ideal distance matrix $D_i$ with values determined by both frequency and co-occurrence. The goal is to map the proportion of intersection area (dark gray in Fig. 1) and circle area to the ratio of frequency and co-occurrence. This is an instance of "area proportional" Euler diagramming [2], which is non-trivial to solve, so we approximate the solution using the distance relation

$$\frac{c_{ij}}{f_i} = \frac{d_o}{r_i}$$

where $c_{ij}$ is the co-occurrence of words $i$ and $j$, $f_i$ is the frequency of word $i$, $r_i$ is the radius of agent $i$, and overlap distance $d_o = D_i[i,j] - r_i - r_j$ is the portion of the ideal distance between agents $i$ and $j$ that overlaps the intersection of their circles. We can then solve for $D_i[i,j]$. Since we can also construct an analogous distance relation for agent $j$, we compromise, filling $D_i$ using the average of the two solutions. When $d_o$ approaches zero, we make $D_i[i,j]$ proportional only to $c_{ij}$. Thus ideally, weakly related words will not intersect, but be located far apart from one another, with completely unrelated words maximally separated.

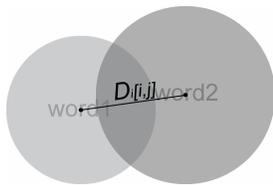

Figure 1: approximating proportional area with distance.

The *scraper* converts online data in RSS, XML or HTML format into a series of *articles*, each of which must contain a body of text and an accompanying list of filtered words extracted from this text. Since a wide number of online sources now include RSS content feeds that can be filtered by API calls, the scraper enables visualization of a wide range of rapidly changing online data streams. Filtering strips out HTML tags and words that are too short, too long, or are in a *stop list* of the commonplace and obscene. Optionally, the scraper can add title and source tag to each article.

The scraper runs in a separate process from the application and visualization infrastructure. This enables it not only to perform parsing and filtering in parallel, but also to support asynchronous handling of user input to filter text streams. This is particularly useful when the interface and display are not co-located.

The *analyzer* measures word frequency across all recent articles. It removes words with extremely low frequency from further analysis. It then creates the co-occurrence matrix (the precursor to $D_i$), where one co-occurrence is counted each time two words appear in the same article. Co-occurrence is a measure of word relatedness.

## 4 RESULTS

Our tool successfully provides users with information about topics of interest within streaming text. The top of Figure 2 shows the results from a search including the word "soldier." As the filter criteria, this word is the most frequently occurring in the resulting stream, and has the largest circle. It also co-occurs at least moderately with all terms, pulling it to the center. There are three larger clusters in this visualization, depicting three distinct stories: one involving a military chase, another a military murder, and the third a military tragedy in France. The red tint of the word "Bragg" (among others) indicates that its position is far from ideal. At the bottom of Figure 2, the user has clicked on a cluster of words to reveal source articles. Since the article itself is an agent, other agents move aside. Note that "Bragg" has taken the opportunity to move across the display to a better position.

Of course, our text visualization must be dynamic, responding to changes in the stream. When the stream is unchanging, agents slow, and approach a steady state. Thus velocity indicates system stability.

## 5 LIMITATIONS AND POSSIBILITIES

Our agent-based infrastructure is only partly utilized; many agent properties (e.g. text size, color, etc.) are not meaningful. Further experimentation with these properties should reveal ways to encode even more information in the visualization.

Control of stream filtering is currently only available through a dialog in the scraper. We are currently making this control available from within the visualization itself, enabling a semantic

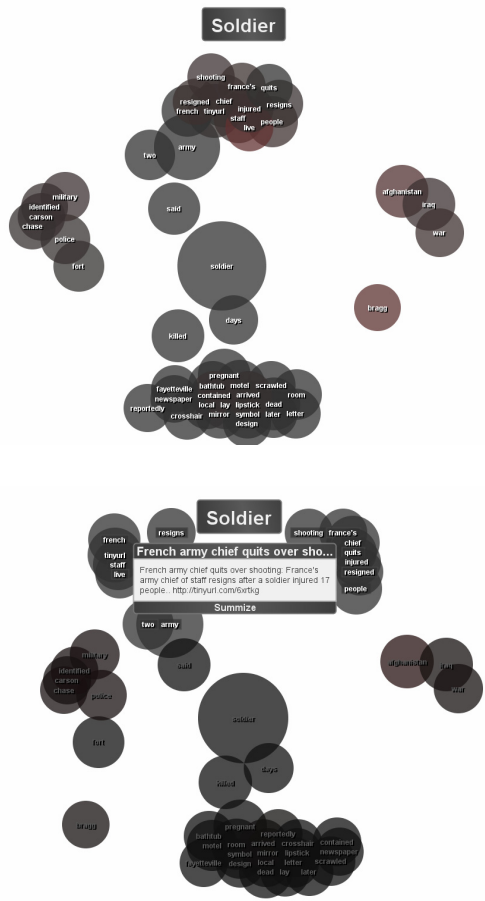

Figure 2: Top, a visualization of a text stream with articles including the word "soldier." Three primary clusters emerge. Bottom, the user examines an article from the top cluster.

zoom by for example clicking on agents, which would issue a new query using the agents' words.

When new data arrives in the stream, our current infrastructure requires a complete refresh of the display, rebuilding from scratch. We are currently implementing an incremental update into the infrastructure.

## ACKNOWLEDGEMENTS

Thanks to Chris Healey for his many thoughts and suggestions.